\documentstyle[12pt]{report}

\begin{document}

\newcommand{\dfrac}[2]{\displaystyle{\frac{#1}{#2}}}

{\it University of Shizuoka}

\hspace*{9.5cm} {\bf US-97-05}\\[-.3in]

\hspace*{9.5cm} {\bf July 1997}\\[.3in]

\vspace*{.4in}

\begin{center}

{\large\bf Neutrino Masses and Mixings}\\[.2in]

{\large\bf in a Universal Seesaw Mass Matrix Model}\\[.3in]

{\bf Yoshio Koide}\footnote{
E-mail: koide@u-shizuoka-ken.ac.jp} \\

Department of Physics, University of Shizuoka \\ 
395 Yada, Shizuoka 422, Japan \\[.1in]

\vspace{.3in}

{\large\bf Abstract}\\[.1in]

\end{center}

\begin{quotation}
Neutrino masses and mixings are investigated on the basis 
of a universal seesaw mass matrix model, in which
quark (except for top) and charged lepton mass matrices
$M_f$  and neutrino mass matrix $M_\nu$ are given by 
$M_f\simeq m_L M_F^{-1} m_R$ and $M_\nu \simeq m_L M_F^{-1} m_L^T$ 
($F=N$), respectively. 
For a simple model which can successfully  describe quark masses 
and mixings, we find that the observed neutrino data 
(except for the solar neutrino data) 
are favor to the intermediate mass scales $O(m_R) = 10^{11}$ GeV 
and $O(M_F)= 10^{13}$ GeV together with $O(m_L)= 10^2$ GeV.
In spite of the largesse of $O(m_R)$, the observed top quark mass 
can be consistently understood from the 
would-be seesaw mass matrix with these mass scales.
\end{quotation}

\vfill
%%%%%%%%%%%%%%%%%%%%%%%%%%%%%%%%%
\newpage

{\large\bf 1. Introduction}
\vglue.05in

One of the most exciting problem in the quark and lepton physics
is to give a natural understanding of the observed hierarchical mass 
spectra of quarks and leptons in relation to the electroweak scale 
$\Lambda_W$.
In the conventional models, the fundamental fermions acquire their masses 
through the symmetry breaking of the electroweak symmetry at the energy 
scale $\mu=\langle \phi_L^0\rangle \equiv \Lambda_W=174$ GeV.
However, the observed mass values except for the top quark mass $m_t$ are 
considerably small compared with $\Lambda_W$. 
Especially, the neutrino masses are invisibly small (or exactly zero).
In relation to this problem, we know an interesting mechanism, the 
so-called ``seesaw" mechanism. 

The  mechanism [1] was first proposed in order to answer 
the question why neutrino masses are so invisibly small: 
$M_\nu \simeq - m M^{-1} m^T$, 
where $m$ is a Dirac mass matrix with the order of the conventional 
quark and charged lepton masses and $M$ is a Majorana mass matrix of 
the right-handed neutrino $\nu_{Ri}$ ($i$ is family-index). 
If we suppose $m_\nu \leq 10$ eV and $m = O(m_\tau)$, the Majorana 
mass $M$ must be larger than $10^8$ GeV.
Usually, in the most models, the order of $M$ is taken as the order 
of the unification energy scale.

On the other hand, in order to understand why quark masses are so small 
compared with the electroweak scale $\Lambda_W$, the mechanism 
was applied to the quarks [2]: $M_f \simeq - m_L M_F^{-1} m_R$, 
where $M_F$ is a mass matrix of hypothetical heavy fermions $F_i$.
If we take $m_L \sim m_W \sim 10^2$ GeV, $m_R\sim 10^3$ GeV and 
$m_f\sim$ a few GeV, then the heavy fermion mass $M_F$ must be of 
the order of $10^5$ GeV.
However, it seems to be not economical that we have two different 
mass scales $M$ and $M_F$. 
Can we build a model with the same mass scale for $M$ and $M_F$?
The answer is Yes.
Since the neutrino and quark mass matrices $M_\nu$ and $M_f$ are 
given by $M_\nu \simeq - m_L M_F^{-1} m_L^T$  ($F=N$) and 
$M_f \simeq - m_L M_F^{-1} m_R$ ($F=U,D$), respectively, we can
understand the smallness of the neutrino masses by assuming 
$O(m_L)/O(m_R)= O(m_\nu/m_f)$ ($f=u,d$) [3].

By the way, for the seesaw model for quarks, there seems to be 
a stumbling block: 
it seems that the observed top quark mass [4]
$m_t\sim \Lambda_W$ apparently takes objection to 
the application of the seesaw mass matrix model to quarks.
However, recently, it has been pointed out [5,6] that
the seesaw mass matrix model for quarks 
is rather preferable to understand why only top quark mass $m_t$ 
is of the order of the electroweak scale $\Lambda_W$ and 
why $m_t$ is so singularly enhanced in the third-family in contrast 
with $m_u\sim m_d$ in the first family.
In the framework of the SU(2)$_L\times$SU(2)$_R\times$U(1)$_Y$ gauge 
model, the $6\times 6$ would-be seesaw mass matrix for the fermion $(f, F)$ 
is given by 
$$
M=\left( 
\begin{array}{cc}
0 & m_L \\
m_R & M_F \\ 
\end{array} \right) 
= m_0\left( 
\begin{array}{cc}
0 & Z_L \\
\kappa Z_R & \lambda Y_f \\ 
\end{array} \right)  \ , 
\eqno(1.1)
$$
where $f_i$ ($i=1,2,3$: family index)  are the ordinary quarks ($f=u,d$) 
and leptons ($f=\nu, e$), and  $F_i$ are hypothetical vector-like 
heavy fermions corresponding to $f_i$.
The fermions $f_i$ and $F_i$ belong to $f_L=(2,1)$, 
$f_R=(1,2)$, $F_L=(1,1)$ and $F_R=(1,1)$ of 
SU(2)$_L \times$SU(2)$_R$, respectively.
The symmetries SU(2)$_L$ and SU(2)$_R$ are broken by the vacuum expectation
values of the Higgs scalars $\phi_L$ and $\phi_R$, i.e., 
the mass matrices $m_L$ and $m_R$, respectively.
For convenience, in (1.1), we have denoted the matrices $m_L$, $m_R$ and 
$M_F$ in terms of the matrices $Z_L$, $Z_R$ and $Y_f$, which are 
of the order of one, i.e., Tr$(Z_L Z_L^\dagger) = O(1)$, 
Tr$(Z_R Z_R^\dagger)= O(1)$ and Tr$(Y_f Y_f^\dagger) = O(1)$.  
It is well-known that the mass matrix (1.1) leads to the seesaw expression  
$$
M_f \simeq - m_L M_F^{-1} m_R \ , \eqno(1.2)
$$
for the case  $|\lambda|\gg |\kappa| \gg 1$ and det$M_F \neq 0$.
On the contrary, for the case det$M_F =0$, the mass spectrum is
given by [5,6]
$$
\begin{array}{cll}
 m_1,\, m_2 & \sim (\kappa/\lambda) m_0 & \ ,  \\
 m_3 & \sim m_0 & = O(m_L) \ ,  \\
 m_4 & \sim \kappa m_0 &  = O(m_R) \ ,  \nonumber \\
 m_5, \, m_6 & \sim \lambda m_0 & = O(M_F) \ ,   \\
\end{array}
\eqno(1.3)
$$
independently of the explicit structures of $Z_L$, $Z_R$ and $Y_f$.
Note that the third fermion mass $m_3^f$ is given by $m_3^f \sim m_0$
without the suppression factor $\kappa/\lambda$.
Therefore, if we build a model such as det$M_F=0$ in up-quark sector,
we can understand why top-quark alone has a mass of the order of $m_L$ 
($\sim \Lambda_W$).
This fact (1.3) was first explicitly demonstrated by Fusaoka and 
the author [5]
on the basis of a special seesaw mass matrix model, where $M_F$ is 
given  by the form [(unit matrix)+(a rank-one matrix)], and then 
re-stressed by Morozumi et al. [6], on the basis of a general study.

Thus, the result (1.3) seems  to support the idea that
the seesaw mass matrix model for the quarks should be taken seriously.
Then, the problem is rather in the neutrino sector: 
Can the seesaw model, which can successfully describe the quark 
masses and mixings, satisfactorily describe the neutrino masses and 
mixings, too?
The purpose of the present paper is to discuss the neutrino 
phenomenology on the bases of an explicit model which can 
satisfactorily describe the quark masses and mixings.

In the next section, we discuss the $12\times 12$ mass matrix of 
the twelve Majorana neutrinos and we will re-derive the well-known 
form $M_\nu \simeq -(m_0/\lambda) Z_L Y_\nu^{-1} Z_L^T$ for the 
conventional light neutrinos.
In Sec.~III, we introduce a model (Model I) with $Z_L=Z_R$ and 
$Y_\nu=$[(unit matrix)+(rank-one matrix)] as a simple example of 
the matrices $Z_L$, $Z_R$ and $Y_F$.
In Sec.~IV, we try to explain all of the present neutrino data, 
the solar, atmospheric and LSND neutrino data, by Model I.
However, we will fail in the simultaneous explanation of the 
three data, and we can give a satisfactory explanation only by 
giving up the explanation of one of the three.
For example, if we give up an explanation of the solar neutrino 
data, we can give the reasonable values of $(\Delta m^2, \sin^2
2\theta)$ for the atmospheric and LSND neutrinos, and 
we obtain the intermediate mass scales $\kappa m_0 \sim 10^{11}$ GeV 
and $\lambda m_0 \sim 10^{13}$ GeV together with $m_0=312$ GeV.
In spite of the largeness of $O(m_R)$, the model will be able to give 
a reasonable value of the top quark mass $m_t$ (and, of course, 
reasonable other quark masses and mixings). 
Finally, Sec.~V is devoted to the conclusions and remarks.

%%%%%%%%%%%%%%%%%%%%%%%%%%%%%%%
\vspace{.2in}

{\large\bf 2. Neutrino Mass Matrix}
\vglue.05in

The neutral lepton mass matrix which is sandwiched between 
$(\overline{\nu}_L, \overline{\nu}_R^c, \overline{N}_L, \overline{N}_R^c)$ 
and $(\nu_L^c, \nu_R, N_L^c, N_R)^T$, where $\nu_L^c \equiv (\nu_L)^c 
\equiv C \overline{\nu}_L^T$ and so on, is given by 
$$
M=\left(
\begin{array}{cccc}
0 & 0 & 0 & m_L \\
0 & 0 & m_R^T & 0 \\
0 & m_R & M_M & M_D \\
m_L^T & 0 & M_D^T & M_M 
\end{array} \right) \ , 
\eqno(2.1)
$$
where $M_D$ ($\equiv M_N$) and $M_M$ are Dirac and Majorana mass 
matrices of the neutral heavy fermions $N_i$.
The matrices $m_L$ and $m_R$ are universal for all fermion sectors, 
i.e., $f=u, d, e, \nu$, and the differences among up-/down- 
quark/lepton masses are generated by the differences of 
$M_F\equiv \lambda Y_f$. 
The heavy fermions $F_i$ belong to $(1,1)$ of 
SU(2)$_L\times$U(2)$_R$.
Besides, the neutral heavy leptons $N_i$ do not have the U(1)-charge.
Therefore, it is likely that when the Dirac masses $(M_D)_{ij}$ 
are generated between $\overline{N}_{Li}$ and $N_{Rj}$, the Majorana 
masses $(M_M)_{ij}$ are also generated between $\overline{N}_{Li}$ and
$N_{Lj}^c$ ($\overline{N}_{Ri}^c$ and $N_{Rj}$) with the same structure 
at the same energy scale $\mu=\lambda m_0$. 
Hereafter, we put $M_M=M_D\equiv M_N\equiv \lambda m_0 Y_\nu$.

For the case $M_M=M_D$, the diagonalization must be done carefully
because the determinant of the $6\times 6$ sub-matrix for 
$(N_L^c, N_R)$ in the $12\times 12$ matrix (2.1) becomes zero.
First, we rotate the matrix (2.1) on the 
$(N_L^c, N_R)$-plane by the angle $\pi/4$.
Then, the mass matrix (2.1) becomes 
$$
M'=\frac{m_0}{\sqrt{2}} \left(
\begin{array}{cccc}
0 & 0 & -Z_L & Z_L \\
0 & 0 & \kappa Z_R^T & \kappa Z_R^T \\
-Z_L^T & \kappa Z_R & 0 & 0 \\
Z_L^T & \kappa Z_R & 0 & 2\sqrt{2}\lambda Y_\nu 
\end{array} \right) \ , 
\eqno(2.2)
$$
We can see that, of the twelve components of the neutrinos, the 
three are approximately described by the mass matrix $2\lambda m_0 Y_\nu$.
The remaining $9\times 9$ mass matrix is given by 
$$
M'\simeq \frac{m_0}{\sqrt{2}} \left(
\begin{array}{ccc}
0 & 0 & -Z_L \\
0 & 0 & \kappa Z_R^T \\
-Z_L^T & \kappa Z_R & 0 \\
\end{array} \right) 
-\frac{m_0}{4\lambda}\left(
\begin{array}{ccc}
Z_L Y^{-1}Z_L^T & \kappa Z_L Y^{-1}Z_R & 0 \\
\kappa Z_R^T Y^{-1} Z_L^T & \kappa^2 Z_R^T Y^{-1} Z_R & 0 \\
0 & 0 & 0
\end{array}\right) \ ,
\eqno(2.3)
$$
where we have used the formula of the seesaw approximation for 
the $(n+m)\times (n+m)$ matrix $M$:
$$
M=\left( 
\begin{array}{cc}
A & B \\
C & D 
\end{array}\right) 
\ \ \Longrightarrow \ \ 
M'\simeq \left(
\begin{array}{cc}
A-BD^{-1}C & 0 \\
0 & D 
\end{array}\right) \ .
\eqno(2.4)
$$
By using 
$$
R\simeq \left(
\begin{array}{ccc}
1 & \varepsilon & 0 \\
-\varepsilon^\dagger & 1 & 0 \\
0 & 0 & 1 
\end{array} \right) \ , \ \ \ 
\varepsilon =\frac{1}{\kappa}Z_L (Z_R^T)^{-1} \ , 
\eqno(2.5)
$$
the matrix (2.3) is transformed into 
\renewcommand{\arraystretch}{2}
$$
RM' R^T \simeq m_0 \left(
\begin{array}{ccc}
-\displaystyle\frac{1}{\lambda} Z_L Y^{-1} Z_L^T & 
-\displaystyle\frac{\kappa}{2\lambda}Z_L Y^{-1} Z_R & 0 \\
-\displaystyle\frac{\kappa}{2\lambda}Z_R^T Y^{-1} Z_L & 
-\displaystyle\frac{\kappa^2}{4\lambda} Z_R^T Y^{-1} Z_R & 
\displaystyle\frac{\kappa}{\sqrt{2}} Z_R \\
0 & \displaystyle\frac{\kappa}{\sqrt{2}} Z_R & 0 
\end{array}\right) \ .
\eqno(2.6)
$$
\renewcommand{\arraystretch}{1}
Therefore, we obtain the following twelve Majorana neutrinos:
(i) three heavy Majorana neutrinos with masses of the order of 
$\lambda m_0$, whose mass matrix is approximately given by 
$$
M_{heavy}\simeq 2 M_N = 2\lambda m_0 Y_\nu \ , 
\eqno(2.7)
$$
(ii) three sets of almost degenerate two Majorana neutrinos 
(the pseudo-Dirac neutrino [7]) with masses of the order of 
$\kappa m_0$, whose mass matrix is approximately given by
\renewcommand{\arraystretch}{2}
$$
M_{PS-D} \simeq \left(
\begin{array}{cc}
-\frac{1}{4}m_R^T M_N^{-1} m_R & \displaystyle\frac{1}{\sqrt{2}} m_R^T \\
\displaystyle\frac{1}{\sqrt{2}} m_R & 0 \\
\end{array} \right) = \kappa m_0 \left(
\begin{array}{cc}
-\displaystyle\frac{\kappa}{4\lambda}Z_R^T Y_\nu^{-1} Z_R & 
\displaystyle\frac{\kappa}{\sqrt{2}} Z_R^T \\
\displaystyle\frac{\kappa}{\sqrt{2}} Z_R & 0 
\end{array} \right) \ ,
\eqno(2.8)
$$
\renewcommand{\arraystretch}{1}
and (iii) three light Majorana neutrinos with masses of the order of 
$(1/\lambda) m_0$,  whose mass matrix is approximately given by
$$
M_\nu \simeq - m_L M_N^{-1} m_L^T = 
-\frac{m_0}{\lambda} Z_L Y_\nu^{-1} Z_L^T \ .
\eqno(2.9)
$$
Note that, differently from the models by Berezhiani and 
Davidson-Wari [3], there are no neutrinos with masses of the order of 
$(\kappa^2/\lambda)m_0$, i.e., whose mass matrix is approximately given by
$m_R M_N^{-1} m_R^T$.

The neutrinos which are described by the mass matrix (2.9) consist of almost 
left-handed neutrinos $\nu_{Li}$. 
Therefore, our task is to seek for such matrix forms of $Z_L$, $Z_R$ 
and $Y_f$ as those can give reasonable quark and lepton masses, 
Cabibbo-Kobayashi-Maskawa (CKM) [8] matrix, and neutrino mixings, 
where the mass matrices of down-quarks and charged leptons are given by 
(1.2), i.e., $M_f \simeq (\kappa/\lambda) Z_L Y_f^{-1} Z_R$ with 
det$Y_f\neq 0$, that of up-quarks is given by (1.1) with det$Y_u=0$, 
and that of neutrinos is given by (2.9).

%%%%%%%%%%%%%%%%%%%%%%%%%%%%%

\vspace{.2in}

{\large\bf 3. A Simple Case}
\vglue.05in

As an explicit model of the universal seesaw mass matrix models which give 
phenomenologically reasonable predictions for quarks, there is 
a model which has been proposed by Fusaoka and the author [5]. 
In their model, the matrices are simply taken by 
$$
Z_L=Z_R\equiv Z \ , \eqno(3.1)
$$
and
$$
Y_f = {\bf 1} + 3 b_f X \ , \eqno(3.2)
$$
where
${\bf 1}$ and $X$ are a $3\times 3$ unit matrix and a rank-one matrix 
normalized as $X^2=X$, respectively.
They have assumed that the matrix $Z$ is given by a diagonal form 
in the family-basis on which $X$ is democratic form:
$$
Z=\left(
\begin{array}{ccc}
z_1 & 0 & 0 \\
0 & z_2 & 0 \\
0 & 0 & z_3 \\
\end{array} \right) \ , \ \ \ 
X=\frac{1}{3}\left(
\begin{array}{ccc}
1 & 1 & 1 \\
1 & 1 & 1 \\
1 & 1 & 1 \\
\end{array} \right) \ . 
\eqno(3.3)
$$
They have used the input values
$$
\frac{z_1}{\sqrt{m_e}}=\frac{z_2}{\sqrt{m_\mu}}=\frac{z_3}{\sqrt{m_\tau}}=
\frac{1}{\sqrt{m_e+m_\mu+m_\tau}} \ ,
\eqno(3.4)
$$
from $M_e\simeq (\kappa/\lambda)m_0 Z\cdot{\bf 1}\cdot Z$ 
by assuming $b_e=0$ in the charged lepton sector as a trial.
Then, for up-quark sector with det$Y_u=0$, i.e., with $b_u=-1/3$, 
they have obtained the successful relations
$$
m_u \simeq \frac{3}{2} \frac{m_e}{m_\tau} \frac{\kappa}{\lambda} m_0 \ , \ \  
m_c \simeq 2 \frac{m_\mu}{m_\tau} \frac{\kappa}{\lambda} m_0 \ , \ \ 
m_t \simeq \frac{1}{\sqrt{3}} m_0 .
\eqno(3.5)
$$
As we noted in (1.3), the third quark mass does not have the suppression 
factor $\kappa/\lambda$.
The ratio of $\kappa/\lambda$ is fixed as $\kappa/\lambda \simeq 0.02$ 
from the observed ratio of $m_c/m_t$.
For down quark sectors, by choosing the complex parameter $b_f$
as $b_d\simeq -e^{i\pi/10}$, they have obtained reasonable results of 
quark masses (not only $m_i^d/m_j^d$, but also $m_i^u/m_j^d$) and 
CKM matrix parameters.
Hereafter, we will refer this model as Model I.

Therefore, our next interest is whether Model I is applicable to 
neutrino phenomenology under the same parameter values or not.
Although they have taken $\kappa=10$ tentatively, the parameter $\kappa$ 
is essentially free because we does not yet observe the right-handed 
weak boson. 
[The case with a small $\kappa$ (e.g., $\sim 10$) is very attractive 
from the phenomenological standpoint, because the case can bring us 
detectable new physics in abundance [9].
However, in the present paper, since we intend to give a unified 
description of quark and neutrino mass matrices without assuming 
further additional intermediate mass scale, we do not consider the case 
with $\kappa\sim 10$.]
Also, although the parameters $b_f$ have been taken as $b_e=0$, 
$b_u=-1/3$ and $b_d\simeq -e^{i\pi/10}$ in Ref.~[5], the value of $b_\nu$ 
in the neutrino sector is still free.
Therefore, for the neutrino masses and mixings, 
we have three adjustable parameters, 
i.e., $b_\nu$ (complex) and $m_0/\lambda$ (real).

For typical values of $b_\nu$, $b_\nu \simeq -1/3$, $b_\nu \simeq -1/2$ 
and $b_\nu \simeq -1$, the masses $m_i^\nu$ ($i=1,2,3$) and 
mixings $U_{\alpha i}$ ($\alpha = e, \mu, \tau$; $i = 1, 2, 3$) for 
three light neutrinos are given as follows [10].

\noindent {\bf Case} $b_\nu = -1/3 + \Delta b_\nu \ \ 
(1 \gg |\Delta b_\nu| \neq 0)$: 
$$
m_1^\nu \simeq \frac{3}{2} \frac{m_e}{m_\tau}\frac{m_0}{\lambda} \ , \ \ 
m_2^\nu \simeq 2 \frac{m_\mu}{m_\tau}\frac{m_0}{\lambda} \ , \ \ 
m_3^\nu \simeq \frac{2\sqrt{2}}{27|\Delta b_\nu|} \frac{m_0}{\lambda} \ , 
\eqno(3.6)
$$
\renewcommand{\arraystretch}{2}
$$
U \simeq \left(\begin{array}{ccc}
1 & \frac{1}{2}\sqrt{\displaystyle\frac{m_e}{m_\mu}} 
& \sqrt{\displaystyle\frac{m_e}{m_\tau}} \\
-\frac{1}{2}\sqrt{\displaystyle\frac{m_e}{m_\mu}} & 1 
& \sqrt{\displaystyle\frac{m_\mu}{m_\tau}} \\
-\frac{1}{2}\sqrt{\displaystyle\frac{m_e}{m_\tau}} 
& -\sqrt{\displaystyle\frac{m_\mu}{m_\tau}} & 1 \\
\end{array} \right)  \ , 
\eqno(3.7)
$$
\renewcommand{\arraystretch}{1}

\noindent {\bf Case} $b_\nu \simeq -1/2$:
$$
m_1^\nu \simeq 2\frac{m_e}{m_\tau}\frac{m_0}{\lambda} \ , \ \ \ m_2^\nu \simeq 
m_3^\nu \simeq \sqrt{\frac{m_\mu}{m_\tau}} \frac{m_0}{\lambda} \ , 
\eqno(3.8)
$$
\renewcommand{\arraystretch}{2}
$$
U\simeq \left(\begin{array}{ccc}
1 & \sqrt{\displaystyle\frac{m_e}{2m_\mu}} 
& \sqrt{\displaystyle\frac{m_e}{2m_\mu}} \\
-\sqrt{\displaystyle\frac{m_e}{m_\mu}} 
& \displaystyle\frac{1}{\sqrt{2}} & 
\mp \displaystyle\frac{1}{\sqrt{2}} \\
-\sqrt{\displaystyle\frac{m_e}{m_\tau}} 
& \pm \displaystyle\frac{1}{\sqrt{2}} & 
\frac{1}{\sqrt{2}} \\
\end{array} \right) \ \ , 
\eqno(3.9)
$$
\renewcommand{\arraystretch}{1}

\noindent {\bf Case} $b_\nu \simeq -1$:
$$
m_1^\nu \simeq m_2^\nu \simeq \sqrt{\frac{m_e m_\mu}{m_\tau^2}}
\frac{m_0}{\lambda} \ , \ \ \ m_3^\nu \simeq 
\frac{1}{2}\frac{m_0}{\lambda} \ , 
\eqno(3.10)
$$
\renewcommand{\arraystretch}{2}
$$
U\simeq \left(\begin{array}{ccc}
\frac{1}{\sqrt{2}} & \mp \displaystyle\frac{1}{\sqrt{2}} 
& -\sqrt{\displaystyle\frac{m_e}{m_\tau}} \\
\pm \displaystyle\frac{1}{\sqrt{2}} 
& \displaystyle\frac{1}{\sqrt{2}} 
& -\sqrt{\displaystyle\frac{m_\mu}{m_\tau}} \\
\sqrt{\displaystyle\frac{m_\mu}{2m_\tau}} 
& \sqrt{\displaystyle\frac{m_\mu}{2m_\tau}} & 
1 \\
\end{array} \right) \ \ . 
\eqno(3.11)
$$
\renewcommand{\arraystretch}{1}

%%%%%%%%%%%%%%%%%%%%%%%%%%%%%%%%%%%%%%%%%%%%%
\vspace{.2in}

{\large\bf 4. Neutrino Data and their Interpretations}
\vglue.05in

As possible evidences for non-zero neutrino masses, at present, 
the following data are known:

\noindent (a) The solar neutrino data [11] with the 
Mikheyev-Smirnov-Wolfenstein (MSW) effect [12] have suggested 
$$
\Delta m_\odot^2 \simeq 6\times 10^{-6} \ {\rm eV}^2 \ , 
\ \ \ \ \ \ \sin^2 2\theta_\odot \simeq 7\times 10^{-3} \ , 
\eqno(4.1)
$$
(the small-angle solution), or 
$$
\Delta m_\odot^2 \simeq  10^{-5} \ {\rm eV}^2 \ , 
\ \ \ \ \ \ \sin^2 2\theta_\odot \simeq 0.8 \ , 
\eqno(4.2)
$$ 
(the large-angle solution).

\noindent (b) The atmospheric neutrino data reported by the 
Kamiokande collaboration [13] have suggested a neutrino mixing 
$\nu_\mu \leftrightarrow \mu_x$: 
$$
\Delta m_{atm}^2 \simeq 1.6 \ (1.8)\times 10^{-2} \ {\rm eV}^2 \ , 
\ \ \ \ \ \sin^2 2\theta_{atm} \simeq 1 \ , 
\eqno(4.3)
$$
for $x=\mu$ ($x=e$).

\noindent (c) The neutrino oscillation 
($\overline{\nu}_\mu\rightarrow \overline{\nu}_e$) experiment 
by the liquid scintillator neutrino detector (LSND) [14] at Los 
Alamos has been reported nonzero neutrino mass: 
$$
(\Delta m^2, \sin^2 2\theta)_{LSND} \simeq 
(0.3\ {\rm eV}^2,\ 0.04)- (2\ {\rm eV}^2,\ 0.002) \ . 
\eqno(4.4)
$$

\noindent (d) A cosmological model with cold+hot dark matter (CHDM)
suggests [15]
$$
m_1^\nu+m^\nu_2+m^\nu_3 \simeq 4.8\ {\rm eV} \ .
\eqno(4.5)
$$
These mass values and mixings (4.1)--(4.3) and (4.5) are not based on 
direct observations of masses and mixings and they are highly 
model-dependent.
On the other hand, the experiment which reported the result (4.4) is 
still controversial [16].
(For convenience, hereafter, we will refer these data (4.1)--(4.5) 
as $\nu_\odot^{small}$, $\nu_\odot^{large}$, $\nu_{atm}$, $\nu_{LSND}$
and $\nu_{CHDM}$, respectively.)

Since $\Delta m^2_{LSND} \gg \Delta m^2_{atm} \gg \Delta m^2_\odot$, 
we investigate only the following two cases: 
(A) $\Delta m_{32}^2 \gg \Delta m_{21}^2$ \ \ 
(i.e., $b_\nu \simeq -1$ and $b_\nu \simeq -1/3$); 
(B) $\Delta m_{32}^2 \ll \Delta m_{21}^2$ \ \ 
(i.e., $b_\nu \simeq -1/2$) \ , where 
$\Delta m_{ij}^2 = m_i^{\nu 2} - m_j^{\nu 2}$. 
Of course, in the three family model, we cannot assign the 
three values $\Delta m_{LSND}^2$, $\Delta m_{atm}^2$ and $\Delta m_\odot^2$ 
to ($\Delta m_{32}^2$, $\Delta m_{21}^2$) simultaneously. 
For the case (A), we consider two cases: 
(A$_1$) $(\Delta m_{32}^2, \ \Delta m_{21}^2) = (\Delta m_{LSND}^2, \ 
\Delta m_\odot^2)$ and (A$_2$) $(\Delta m_{32}^2, \ \Delta m_{21}^2) = 
(\Delta m_{LSND}^2, \ \Delta m_{atm}^2)$. 
Similarly, for the case (B), we consider two cases: 
(B$_1$) $(\Delta m_{32}^2, \Delta m_{21}^2) =$ 
$(\Delta m_\odot^2, \ \Delta_{LSND}^2)$ and $(\Delta m_{32}^2, 
\Delta m_{21}^2) =$ $(\Delta m_{atm}^2, \Delta m_{LSND}^2)$. 
We do not consider the cases $(\Delta m_{atm}^2, \ 
\Delta m_\odot^2)$ and $(\Delta m_\odot^2, \ \Delta m_{atm}^2)$,
because it is readily known that the cases cannot give the 
observed large mixing $\sin^2 2\theta_{atm}\sim 1$ in $\nu_{atm}$. 

In the three-family model, the neutrino oscillation $P(\nu_\alpha 
\rightarrow \nu_\beta) \ (\alpha \neq \beta)$ is given by 
$$
P(\nu_\alpha \rightarrow \nu_\beta) = S_{21}^{\alpha \beta} S_{21} + 
S_{31}^{\alpha \beta} S_{31} + S_{32}^{\alpha \beta} S_{32} \ \ , 
\eqno(4.6)
$$
where 
$$
S_{ij}^{\alpha \beta} = -4 \Re (U_{\alpha i} U_{\beta i}^* 
U_{\alpha j}^* U_{\beta j}) \ \ , 
\eqno(4.7)
$$
$$
S_{ij} = \sin^2 (L\Delta m_{ij}^2/4 E_\nu) \ \ , 
\eqno(4.8)
$$
and we have neglected $CP$-violation terms in (4.6).

In the case (A$_1$), since $\Delta m_{32}^2 \gg 
\Delta m_{atm}^2 \gg \Delta m_{21}^2$, 
we can regard $S_{ij}$ as $S_{21} \simeq 0$ and 
$\langle S_{31}\rangle = \langle S_{32}\rangle = 1/2$ 
for the atmospheric neutrinos, where $\langle S_{ij}\rangle$
denotes the mean value of $S_{ij}(L/E_\nu)$. 
Then, by using $P(\nu_\alpha \rightarrow \nu_\beta) \simeq 
2|U_{\alpha 3}|^2 |U_{\beta 3}|^2$, the ratio 
$$
R_{atm} = \frac{(\nu_\mu/\nu_e)_{data}}{(\nu_\mu/\nu_e)_{MC}} \simeq 
\frac{P(\nu_\mu \rightarrow \nu_\mu) + \frac{1}{2} P(\nu_e \rightarrow 
\nu_\mu)}{P(\nu_e \rightarrow \nu_e) + 2P(\nu_\mu \rightarrow \nu_e)} \ \ , 
\eqno(4.9)
$$
is expressed as 
$$
R_{atm} \simeq \frac{1 - |U_{e3}|^2 |U_{\mu 3}|^2 - 2 |U_{\mu 3}|^2 
|U_{\tau 3}|^2}{1 + 2|U_{e3}|^2|U_{\mu 3}|^2 - 
2|U_{e3}|^2|U_{\tau 3}|^2} \ \ . 
\eqno(4.10)
$$
Therefore, in the case (A$_1$), instead of (4.3), we temporize 
with (4.10) whose value is consistent with the observed value [13] 
$$
R_{atm} = 0.57^{+0.08}_{-0.07} \pm 0.07 \ \ . 
\eqno(4.11)
$$
However, both cases $b_\nu\simeq -1/3$ and $b_\nu\simeq -1$, 
from (3.7) and (3.11), respectively, give 
$R_{atm}\simeq 1-2m_\mu/m_\tau \simeq 1$, 
which is in disagreement with the observed value (4.11).
On the other hand, the mixing parameter $\sin^2 2\theta_\odot$ in the 
MSW solutions is interpreted by 
$$
\sin^2 2\theta_\odot = 4|U_{e1}|^2|U_{e2}|^2 \ \ . 
\eqno(4.12)
$$
The cases $b_\nu\simeq -1/3$ and $b_\nu\simeq -1$ give
$$
\sin^2 2\theta_\odot \simeq m_e/m_\mu \simeq 0.005 \ , 
\eqno(4.13)
$$
and
$$
\sin^2 2\theta_\odot \simeq 1 \ ,
\eqno(4.14)
$$
which are consistent with the small- and large-angle solutions 
(4.1) and (4.2), respectively.

In the case (A$_2$), since 
$\Delta m^2_\odot \ll \Delta m^2_{21} \ll \Delta m^2_{32}$, 
we can regard $S_{ij}$ as $\langle S_{21}\rangle =\langle S_{31}\rangle 
=\langle S_{32}\rangle =1/2$ for $\nu_\odot$.
Then, we obtain
$$
P_{ee}\equiv P(\nu_e \rightarrow \nu_e)=
|U_{e1}|^4+|U_{e2}|^4+|U_{e3}|^4 \ .
\eqno(4.15)
$$
According to the recent analysis by Acker and Pakvasa [17] 
we will search for a solution which is consistent with  
$$
P_{ee} = 0.42 - 0.52 \ . 
\eqno(4.16)
$$
The cases $b_\nu\simeq -1/3$ and $b_\nu\simeq -1$, from (3.7) and 
(3.11), give $P_{ee}\simeq 1$ and $P_{ee}\simeq 1/2$, 
respectively.
Therefore, only the case $b_\nu\simeq -1$ can explain the solar 
neutrino data.
On the other hand, the mixing parameter $\sin^2 2 \theta_{atm}$ 
is interpreted by 
$$
\sin^2 2 \theta_{atm} = S_{21}^{\mu e} = 
-4 \Re(U_{\mu 2}U_{e 2}^* U_{\mu 1}^* U_{e 1}) \ . 
\eqno(4.17)
$$
The cases $b_\nu\simeq -1/3$ and $b_\nu\simeq -1$ give
$\sin^2 2 \theta_{atm} \simeq m_e/m_\mu \simeq 0.005$ and
$\sin^2 2 \theta_{atm} \simeq 1$, respectively.
Again, only the case $b_\nu\simeq -1$ is favorable to the data.

In both cases (A$_1$) and (A$_2$), the mixing parameter 
$\sin^2 2 \theta _{LSND}$  is given by 
$$
\sin^2 2 \theta_{LSND} = 4|U_{e 3}|^2 |U_{\mu 3}|^2  \ \ . 
\eqno(4.18)
$$
The cases $b_\nu\simeq -1/3$ and $b_\nu\simeq -1$ give
$\sin^2 2 \theta_{LSND}\simeq 4 m_e m_\mu/m_\tau^2 \simeq 7\times 
10^{-5}$, which is too small compared with the observed mixing 
value (4.4).

Similarly, for the case (B$_1$), we use the expressions  
$$
\sin^2 2 \theta_\odot = 4|U_{e 3}|^2 |U_{e 2}|^2 \ \ . 
\eqno(4.19)
$$
$$
R_{atm} \simeq \frac{1 - |U_{e1}|^2 |U_{\mu 1}|^2 - 2 |U_{\mu 1}|^2 
|U_{e 1}|^2}{1 + 2|U_{e1}|^2|U_{\mu 1}|^2 - 
2|U_{e1}|^2|U_{\tau 1}|^2} \ \ . 
\eqno(4.20)
$$
The mixing matrix (3.9) gives 
$\sin^2 2 \theta_\odot \simeq (m_e/m_\mu)^2 \simeq 2\times 10^{-5}$ 
and $R_{atm} \simeq 1-m_e/2m_\mu \simeq 1$, 
which are in disagreement with the data.

In the case (B$_2$), the solar neutrino data is explained by $P_{ee}$ 
given by (4.15), and the mixing parameter in $\nu_{atm}$ is 
expressed by 
$$
\sin^2 2 \theta_{atm} = S_{23}^{\mu \tau} = -4 Re (U_{\mu 2}
U_{\tau 2}^* U_{\mu 3}^* U_{\tau 3}) \ . 
\eqno(4.21)
$$
Since the mixing matrix (3.9) gives $P_{ee}\simeq 1$ and 
$\sin^2 2 \theta_{atm}\simeq 1$, 
we fail to explain the $\nu_\odot$ data, but we can understand
the large mixing in $\nu_{atm}$.

For both cases (B$_1$) and (B$_2$), the mixing parameter 
$\sin^2 2 \theta_{LSND}$ is given by 
$$
\sin^2 2 \theta_{LSND} = 4 |U_{e1}|^2 |U_{\mu 1}|^2
\simeq 4m_e/m_\mu \simeq 0.02 \ , 
\eqno(4.22)
$$
where we have used the mixing matrix (3.9) for $b_\nu\simeq -1/2$.
The mixing value (4.22) corresponds to $\Delta m^2_{LSND}\simeq 0.5$
eV$^2$ in the LSND allowed region (4.4).

The results are summarized in Table I. We find that the present 
model cannot give a simultaneous interpretation for the three neutrino 
data, $\nu_\odot$, $\nu_{atm}$ and $\nu_{LSND}$. 
We must give up the explanation of one of the three data. 

As suggested by the case (B$_2$) in Table I, if we give up explaining 
the solar neutrino data, we can find an interesting solution: 
$(\Delta m_{32}^2, \ \Delta m_{21}^2) = (\Delta m_{atm}^2, \ 
\Delta m_{LSND}^2)$ with $b_\nu \simeq -1/2$. 
For example, the solution $b_\nu = -(1/2) e^{i \beta_\nu}$ with 
$b_\nu = 0.12^\circ$ gives 
$$
\sin^2 2 \theta_{atm} = 0.995 \ , \ \ \sin^2 2 \theta_{LSND} = 0.0191 \ \ , 
\eqno(4.23)
$$
$$
\begin{array}{l}
m_1^\nu = 0.000540\ m_0/\lambda \ ,  \\
m_2^\nu = 0.2288\ m_0/\lambda \ ,  \\
m_3^\nu = 0.2326\ m_0/\lambda \ ,  \\
\end{array}
\eqno(4.24)
$$
$$
r \equiv \Delta m_{32}^2 / \Delta m_{21}^2 = 0.0331 \ \ . 
\eqno(4.25)
$$
The input value $\Delta m_{32}^2 = \Delta m_{atm}^2 = 0.016$ ${\rm eV}^2$ 
predicts 
$$
m_1^\nu = 0.0016\ {\rm eV} \ , \ \ 
m_2^\nu = 0.695\ {\rm eV} \ , \ \ 
m_3^\nu = 0.707\ {\rm eV} \ , 
\eqno(4.26)
$$
$$
\Delta m_{21}^2 = 0.483\ {\rm eV}^2 \ , 
\eqno(4.27)
$$
together with
$$
m_0/\lambda = 3.04\ {\rm eV} \ . 
\eqno(4.28)
$$
The values (4.23) and (4.27) are in good agreement with the 
observed SLND solution $(\Delta m^2, \ \sin^2 2 \theta)_{LSND} \simeq 
(0.5\ {\rm eV}^2, \ 0.02)$. 
Although the sum $\sum m_i^\nu = 1.4$ ${\rm eV}$ is somewhat small 
compared with the mass value in the CHDM scenario (4.5), 
these neutrinos can still be one of the dark matter objects

When the LSND data are neglected, the case (A$_2$) with
$b_\nu\simeq -1$ is also interesting.
For example, the parameter value $b_\nu=-e^{i\beta_\nu}$ 
with $\beta_\nu=1.8^\circ$ gives
$$
P_{ee}=0.52 \ , \ \ \ \sin^2 2\theta_{atm}=0.90 \ ,
\eqno(4.29)
$$
which are favorable to the data of $\nu_\odot$ and $\nu_{atm}$. 
On the other hand, the input
$$
\Delta m^2_{21}=1.283\times 10^{-5} (m_0/\lambda)^2
=\Delta m^2_{atm}=1.8\times 10^{-2} \ {\rm eV}^2 \ , 
\eqno(4.30)
$$
leads to 
\setcounter{chapter}{4}
\setcounter{equation}{30}
\begin{eqnarray}
m_1^\nu & = & 0.00305\ m_0/\lambda = 0.114\ {\rm eV} \ , \nonumber \\
m_2^\nu & = & 0.00471\ m_0/\lambda = 0.176\ {\rm eV} \ , \nonumber \\
m_3^\nu & = & 0.5002\ m_0/\lambda  = 18.7 \ {\rm eV} \ , \nonumber \\
\end{eqnarray}
together with
$$
m_0/\lambda =37.5 \ {\rm eV} \ .
\eqno(4.32)
$$
The value $\sum m_i^\nu \simeq 19$ eV is somewhat large to 
identify these neutrinos as those in the dark matter scenario.
In the present case, although the third neutrino has a considerably 
large value $m_3^\nu=18.7$ eV, the effective electron neutrino 
mass $\langle m_\nu \rangle$ is safely small compared with the 
upper bound on $\langle m_\nu \rangle$ from the 
neutrinoless double beta decays, $\langle m_\nu \rangle <0.68$ eV
[18], because of the smallness of $|U_{e3}^2|$, i.e., 
$$
\langle m_\nu \rangle =\left| m_1^\nu U_{e1}^2+m_2^\nu U_{e2}^2
+m_3^\nu U_{e3}^2 \right| = 0.0027\ {\rm eV} \ .
\eqno(4.33)
$$ 

In these examples, we have chosen the solutions with 
$\beta_\nu\neq 0$.
However, this is not essential for the numerical results. 
The similar results to the cases $b_\nu=-(1/2)\exp[i\, 0.12^\circ]$
and  $b_\nu=-\exp[i\, 1.8^\circ]$ can also be obtained by 
choosing $b_\nu=-0.50004$ and $b_\nu=-0.97$. 
However, since we have taken the rational solutions of $|b_f|$ 
in the quark and charged lepton sectors, i.e., 
$|b_u|=1/3$, $|b_d|=1$ and $|b_e|=0$ [5],
we have taken the rational solutions of $|b_\nu|$ in Table II.

In Table II, we summarize these typical cases.
For reference, in Table II, we show the values of 
$\sin^2 2\theta^{\mu\tau}$ which is given by 
$\sin^2 2\theta^{\mu\tau}= 4|U_{\mu 3}|^2 |U_{\tau 3}|^2$ 
for the case $\Delta m^2_{32}\gg \Delta m^2_{21}$ and by 
$\sin^2 2\theta^{\mu\tau}= 4|U_{\mu 1}|^2 |U_{\tau 1}|^2$ 
for the case $\Delta m^2_{32}\ll \Delta m^2_{21}$, 
because CHORUS [19] and NOMAD [20] experiments at CERN are expected 
to present results on the $\nu_\mu \rightarrow \nu_\tau$ oscillation 
in the very near future.
As pointed out by Tanimoto [21], the values of 
$\sin^2 2\theta^{\mu\tau}$ are very small for the case 
$m_1^\nu \ll m_2^\nu \simeq m_3^\nu$, i.e., 
$\sin^2 2\theta^{\mu\tau}\simeq 4m_e/m_\tau$ for $b_\nu\simeq -1/2$,
so that it will be hopeless to observe the 
$\nu_\mu \rightarrow \nu_\tau$ oscillation at CHORUS and NOMAD.
The case $\Delta m^2_{32}\gg \Delta m^2_{21}$, i.e., 
$b_\nu\simeq -1$, gives a sizable value 
$\sin^2 2\theta^{\mu\tau}\simeq 4m_\mu/m_\tau$, 
so that the case will be hopeful to 
observe the evidence of the $\nu_\mu \rightarrow \nu_\tau$ 
oscillation.

In Table II, we also list the value of the intermediate mass 
scales $\kappa m_0$ and $\lambda m_0$, which are obtained by 
the following relations:
$$
m_t\simeq \frac{1}{\sqrt{3}}  m_0 \simeq 180\  {\rm GeV} \ \ 
\ \  ({\rm at}\ \mu=m_Z) \ , 
\eqno(4.34)
$$
from (3.5) and  
$$
(\kappa/\lambda) m_0 = m_\tau + m_\mu +m_e =1.850 \ {\rm GeV} \ ,
\ \  ({\rm at}\ \mu=m_Z)
\eqno(4.35)
$$
from  $M_e\simeq (\kappa/\lambda)m_0 Z\cdot {\bf 1}\cdot Z$, i.e., 
$$
m_0\simeq \sqrt{3}  m_t \simeq 312\  {\rm GeV} \ \ 
\ \  ({\rm at}\ \mu=m_Z) \ , 
\eqno(4.36)
$$
$$
\kappa/\lambda = 5.93 \times 10^{-3} \ .
\eqno(4.37)
$$

%%%%%%%%%%%%%%%%%%%%%%%%%%%%%%%%%%%%%%

\vspace{.2in}
{\large\bf 5. Conclusions}
\vglue.05in

In conclusion, we have investigated a universal seesaw mass matrix 
model with a form (1.1), and have found that the model has a 
possibility of giving a unified description of a mass hierarchy of 
quarks and leptons by assuming det$M_F=0$ for the heavy fermion 
mass matrix $M_F$ in the up-quark sector.
Here, the ``hierarchy" discussed in the present paper means 
the mass hierarchy among the three groups, (i) top quark, (ii) 
quarks except for top and  charged leptons, and (iii) neutrinos.
For example, in Model I, the hierarchy among $m_e$, $m_\mu$ and 
$m_\tau$ are given by hand, i.e., by adjusting the parameters $z_i$ 
in the order-one matrices $Z\equiv Z_L=Z_R$.
It is our task at the next step to answer why these 
matrices $Z_L$, $Z_R$ and $Y_f$ take such structures.

In the SU(2)$_L\times$SU(2)$_R\times$U(1)$_Y$ model, it is likely 
that the neutral heavy leptons acquire Majorana masses $M_M$ 
together with the Dirac masses $M_N$ with the same structure and 
the same magnitude, i.e., $M_M=M_N$.
Then, masses of the twelve Majorana neutrinos are 
given by (2.7)--(2.9).
In the present model, heavy neutrinos with masses of the order 
of $(\kappa^2/\lambda)m_0$ do not appear, and, instead, 
pseudo-Dirac neutrinos with masses of the order of $\kappa m_0$
appear.

A suitable choice of the matrices $Z_L$, $Z_R$ and $Y_f$ will 
give a unified descriptions of masses and mixings of quarks 
and leptons.
As a simple example of $Z_L$, $Z_R$ and $Y_f$, a model with 
(3.1)--(3.3), Model I,  has been investigated.
The model can give a unified description of masses and mixings 
of quarks and charged leptons [5]. 
However, the straightforward application  to the neutrino 
phenomenology fails to give the simultaneous explanation 
of the solar, atmospheric and LSND neutrino data.
If we give up explaining one of these three data, for example, 
if we give up the explanation of the solar neutrino data, 
we obtain the intermediate mass scales $\kappa m_0 \sim 10^{11}$ GeV 
and $\lambda m_0 \sim 10^{13}$ GeV together with $m_0=312$ GeV.
Note that in spite of the largeness of $O(m_R)$, the model can give 
a reasonable value of the top quark mass $m_t$. 
(In other words, the above intermediate mass scales have 
estimated from the input value $m_t(m_Z)=180$ GeV.)

If we take all of the neutrino data, $\nu_\odot$, $\nu_{atm}$ 
and $\nu_{LSND}$ seriously, 
we must seek another set of the matrices 
$Z_L$, $Z_R$ and $Y_f$ or 
we must abandon the idea $O(M_N) = O(M_F)$ $(F = E, U, D)$.
Although the model with $Z_L=Z_R$, Model I, has successfully 
described the quark masses and their mixings in terms of 
charged lepton masses, there is no reason that we should 
consider $Z_L=Z_R$.
It will be worth while investigating a model with 
$Z_L\neq Z_R$ for quarks and leptons.
It will also be worth while investigating a model with 
$O(M_N)\gg O(M_F)$ ($F=E, U, D$). 
The latter possibility will bring abundant light neutrinos into 
the model (for example, see a model given in [10]), although 
it is not economical because it bring additional intermediate 
mass scale into the model.

Model I discussed in Secs.~III and IV is only an example. 
There will be many interesting versions of 
the universal seesaw model.
The universal seesaw model will be one of the most 
promising models of the quark and lepton unification.

%%%%%%%%%%%%%%%%%%%%%%%%%%%%%%%%%%%%%%
\vglue.2in

\centerline{\large\bf Acknowledgments}

The author would like to thank M.~Tanimoto for his helpful comments 
and calling his attention to CHORUS and NOMAD experiments. 
He also thank N.~Muto for numerical check of the approximate 
expressions (2.7)--(2.9) of the $12\times 12$ would-be seesaw mass 
matrix (2.1) by his computer program with high accuracy.
This work was supported by the Grant-in-Aid for Scientific Research, the 
Ministry of Education, Science and Culture, Japan (No.08640386). 

%%%%%%%%%%%%%%%%%%%%%%%%%%%%%%%%%%%%%%%%%%%%%%%%%%%%%%%%%%%%%%%%%%%%%%%%%%%%
%%%%%%%%%%%%%%%%
\vglue.2in
%\newpage

\newcounter{0000}
\centerline{\large\bf References}
\begin{list}
{[~\arabic{0000}~]}{\usecounter{0000}
\labelwidth=0.8cm\labelsep=.1cm\setlength{\leftmargin=0.7cm}
{\rightmargin=.2cm}}
\item M.~Gell-Mann, P.~Rammond and R.~Slansky, in {\it Supergravity}, 
edited by P.~van Nieuwenhuizen and D.~Z.~Freedman (North-Holland, 
1979); 
T.~Yanagida, in {\it Proc.~Workshop of the Unified Theory and 
Baryon Number in the Universe}, edited by A.~Sawada and A.~Sugamoto 
(KEK, 1979); 
R.~Mohapatra and G.~Senjanovic, Phys.~Rev.~Lett.~{\bf 44}, 912 (1980).
\item Z.~G.~Berezhiani, Phys.~Lett.~{\bf 129B}, 99 (1983);
Phys.~Lett.~{\bf 150B}, 177 (1985);
D.~Chang and R.~N.~Mohapatra, Phys.~Rev.~Lett.~{\bf 58},1600 (1987); 
A.~Davidson and K.~C.~Wali, Phys.~Rev.~Lett.~{\bf 59}, 393 (1987);
S.~Rajpoot, Mod.~Phys.~Lett. {\bf A2}, 307 (1987); 
Phys.~Lett.~{\bf 191B}, 122 (1987); Phys.~Rev.~{\bf D36}, 1479 (1987);
K.~B.~Babu and R.~N.~Mohapatra, Phys.~Rev.~Lett.~{\bf 62}, 1079 (1989); 
Phys.~Rev. {\bf D41}, 1286 (1990); 
S.~Ranfone, Phys.~Rev.~{\bf D42}, 3819 (1990); 
A.~Davidson, S.~Ranfone and K.~C.~Wali, 
Phys.~Rev.~{\bf D41}, 208 (1990); 
I.~Sogami and T.~Shinohara, Prog.~Theor.~Phys.~{\bf 66}, 1031 (1991);
Phys.~Rev. {\bf D47}, 2905 (1993); 
Z.~G.~Berezhiani and R.~Rattazzi, Phys.~Lett.~{\bf B279}, 124 (1992);
P.~Cho, Phys.~Rev. {\bf D48}, 5331 (1994); 
A.~Davidson, L.~Michel, M.~L,~Sage and  K.~C.~Wali, 
Phys.~Rev.~{\bf D49}, 1378 (1994); 
W.~A.~Ponce, A.~Zepeda and R.~G.~Lozano, 
Phys.~Rev.~{\bf D49}, 4954 (1994).
\item Z.~G.~Berezhiani, in Ref.~[2]; A.~Davidson and K.~C.~Wali, 
in Ref.~[2]; S.~Rajpoot, in Ref.~[2]; A.~Davidson, S.~Ranfone and 
K.~C.~Wali, in Ref.~[2]; W.~A.~Ponce, A.~Zepeda and R.~G.~Lozano, 
in Ref.~[2].
\item CDF collaboration, F.~Abe {\it et al}., Phys.~Rev.~Lett. 
{\bf 73}, 225 (1994).
\item Y.~Koide and H.~Fusaoka, Z.~Phys. {\bf C71}, 459 (1966);
Prog.~Theor.~Phys. {\bf 97}, 459 (1997).
\item T.~Morozumi, T.~Satou, M.~N.~Rebelo and M.~Tanimoto, 
Preprint HUPD-9704 (1977), hep-ph/9703249.
\item L.~Wolfenstein, Nucl.~Phys. {\bf B185}, 147 (1981); 
S.~T.~Petcov, Phys.~Lett. {\bf 110B}, 245 (1982);
C.~N.~Leung and S.~T.~Petcov, Nucl.~Phys. {\bf 125B}, 461 (1983); 
M.~Doi, M.~Kenmoku, T.~Kotani, H.~Nishiura and E.~Takasugi, 
Prog.~Theor.~Phys. {\bf 70}, 1331 (1983):
J.~W.~F.~Valle, Phys.~Rev. {\bf D27}, 1672 (1983);
J.~W.~F.~Valle and M.~Singer, Phys.~Rev. {\bf D28}, 540 (1983);
D.~Wyler and L.~Wolfenstein, Nucl.~Phys. {\bf B218}, 205 (1983).
\item N.~Cabibbo, Phys.~Rev.~Lett.~{\bf 10}, 531 (1996); 
M.~Kobayashi and T.~Maskawa, Prog.~Theor.~Phys.~{\bf 49}, 652 (1973).
\item Y.~Koide,  Preprint US-96-09 (1996), hep-ph/9701261 (unpublished);
US-97-04 (1997), hep-ph/9706277, to be published in Phys.~Rev. 
{\bf D56}, (1997).
\item Y.~Koide,  Mod.~Phys.~Lett.{\bf A36}, 2849 (1996).
\item GALLEX collaboration, P.~Anselmann {\it et al}, 
Phys.~Lett. {\bf B327}, 377 (1994); {\bf B357}, 237 (1995);
SAGE collaboration, J.~N.~Abdurashitov {\it et al}, 
Phys.~Lett. {\bf B328}, 234 (1994). 
Also see, N.~Hata and P.~Langacker, Phys.~Rev. {\bf D50}, 632 (1994); 
{\bf D52}, 420 (1995).
\item S.~P.~Mikheyev and A.~Yu.~Smirnov, Yad.~Fiz. {\bf 42}, 1441 (1985); 
[Sov.~J.~Nucl.~Phys. {\bf 42}, 913 (1985)]; 
Prog.~Part.~Nucl.~Phys. {\bf 23}, 41 (1989); 
L.~Wolfenstein, Phys.~Rev. {\bf D17}, 2369 (1978); {\bf D20}, 2634 (1979);
T.~K.~Kuo and J.~Pantaleon, Rev.~Mod.~Phys. {\bf 61}, 937 (1989). 
Also see, A.~Yu.~Smirnov, D.~N.~Spergel and J.~N.~Bahcall, 
Phys.~Rev. {\bf D49}, 1389 (1994).
\item Y.~Fukuda {\it et al.}, Phys.~Lett. {\bf B335}, 237 (1994).
Also see, Soudan-2 collaboration, M.Goodman {\it et al.}, 
Nucl.~Phys. (Proc.~Suppl.) {\bf B38}, 337 (1995); 
IMB collaboration, D.~Casper {\it  et al}, Phys.~Rev.~Lett. {\bf 66}, 2561 
(1989); R.~Becker-Szendy {\it et al}, Phys.~Rev. {\bf D46}, 3720 (1989).
%
%\item NUSEX collaboration, M.~Aglietta {\it at al.}, Europhys.~Lett. {\bf 8}, 
%611 (1989);
%Frejus collaboration, Ch.~Berger {\it el al.}, Phys.~Lett. {\bf B227}, 489 
%(1989); {\it ibid} {\bf B245}, 305 (1990); K. Daum {\it et al.}, Z.~Phys. 
%{\bf C66}, 417 (1995).
%
\item C.~Athanassopoulos {\it et al.}, Phys.~Rev.~Lett. {\bf 75}, 2650 
(1995);  Phys.~Rev.~Lett. {\bf 77}, 3082 (1996); nucl-ex/9706006 (1997).
\item J.~R.~Primack, J.~Holtzman, A.~Klypin and D.~O.~Caldwell,
Phys.~Rev.~Lett. {\rm 74}, 2160 (1995); 
D.~Pogosyan and A.Starobinsky, astro-ph/9502019.
\item J.~E.~Hill, Phys.~Rev.~Lett. {\bf 75}, 2654 (1995).
%
%\item D.~O.~Caldwell and R.~N.Mohapatra, Phys.~Rev. {\bf D48}, 3259 
%(1993);
%A.~S.~Joshipura, Z.~Phys. {\bf C64}, 31 (1994); 
%S.~T.~Petcov and A.~Smirnov, Phys.~Lett. {\bf B322}, 109 (1994).
%
\item A.~Acker and S.~Pakvasa, Phys.~Lett. {\bf B397}, 209 (1997).
\item M.~K.~Moe, Nucl.~Phys. (Proc.~Suppl.) {\bf B38}, 36 (1995):
A.~Balysh {\it et al.}, Proc. of the Inst. Conference on High 
Energy Physics (ICHEP), 27th, Glashgow, Scotland, Eds, 
P.~J.~Bussey and I.~G.~Knowles, IOP, 1995, p.939 (hep-ex/9502007).
\item M.~Baldo-Ceolin, Nucl.~Phys. (Proc.~Suppl.) {\bf 35}, 450 (1994); 
K.~Winter, {\it ibid}. {\bf 38}, 211 (1995). 
\item L.~DiLella, Nucl.~Phys. (Proc.~Suppl.) {\bf 31}, 319 (1993).
\item M.~Tanimoto, Phys.~Rev. {\bf D53}, 6632 (1996).
\end{list}

\newpage
%%%%%%%%%%%%%%%%%%%%%%%%%%%%%%%%%%%%%%%%%%%%%%%%%
\begin{quotation}
Table I. Possible interpretations for the $\nu_\odot$, $\nu_{atm}$ and 
$\nu_{LSND}$ data. 
The parameter $r$ denotes $r = \Delta m_{32}^2/\Delta m_{21}^2$. 
The numerical results have obtained from the direct evaluation 
of the mass matrix (2.1) without using the approximate expressions 
(3.7), (3.9) and (3.11).
The values with the symbols $\circ$ and $\bullet$ denote that 
those are favorable 
and unfavorable to the observed data, respectively. 
\end{quotation}

\begin{center}
\begin{tabular}{|l|l|l|}\hline\hline
\multicolumn{2}{|c|}{ (A) $\Delta m_{32}^2 \gg \Delta m_{21}^2$} & 
(B) $\Delta m_{32}^2 \ll \Delta m_{21}^2$ \\ \hline
\ \ \ \ $b_\nu \simeq -1$ & \ \ \  \ $b_\nu \simeq -1/3$ &  
\ \ \ \ $b_\nu \simeq -1/2$ \\ \hline
\multicolumn{2}{|c|}{ (A$_1$) $(\Delta m_{32}^2, \Delta m_{21}^2) 
= (\Delta m_{LSND}^2, \Delta m_\odot^2)$} & 
(B$_1$) \ \ $= (\Delta m_\odot^2, 
\Delta m_{LSND}^2)$ \\ \hline
$\circ$ $\sin^2 2 \theta_\odot \simeq 1$ & 
$\circ$ $\sin^2 2 \theta_\odot \simeq 0.04$ & 
$\bullet$ $\sin^2 2 \theta_\odot \simeq 2 \times 10^{-5}$ \\ 
$\bullet$ $R_{atm} \simeq 0.9$ & 
$\bullet$ $R_{atm} \simeq 1$ & 
$\bullet$ $R_{atm} \simeq 1$ \\
$\bullet$ $\sin^2 2\theta_{LSND} \simeq 5 \times 10^{-5}$ & 
$\bullet$ $\sin^2 2 \theta_{LSND} \simeq 6 \times 10^{-5}$ & 
$\circ$ $\sin^2 2 \theta_{LSND} \simeq 0.02$ \\
$\circ$ $r \geq 10^4$ & $\circ$ $r \geq 10^4$ & 
$\circ$ $r \leq 10^{-3}$ \\ \hline
\multicolumn{2}{|c|}{ (A$_2$) $(\Delta m_{32}^2, \Delta m_{21}^2) = 
(\Delta m_{LSND}^2, \Delta m_{atm}^2)$} & 
(B$_2$) \ \ $= (\Delta m_{atm}^2, \Delta m_{LSND}^2)$ 
\\ \hline
$\circ$ $P_{ee} \simeq 0.5$ & 
$\bullet$ $P_{ee} \simeq 1$ & $\bullet$ $P_{ee} \simeq 1$ \\
$\circ$ $\sin^2 2 \theta_{atm} \simeq 1$ & 
$\bullet$ $\sin^2 2 \theta_{atm} \simeq 4 \times 10^{-3}$ & 
$\circ$ $\sin^2 2 \theta_{atm} \simeq 1$ \\
$\bullet$ $\sin^2 2 \theta_{LSND} \simeq 5 \times 10^{-5}$ & 
$\bullet$ $\sin^2 2 \theta_{LSND} \simeq 6 \times 10^{-5}$ & 
$\circ$ $\sin^2 2 \theta_{LSND}\simeq 0.02$ \\
$\circ$ $r \geq 10^4$ & $\circ$ $r \geq 10^4$ & 
$\circ$ $r \leq 10^{-3}$ \\ \hline\hline
\end{tabular}
\end{center}

%%%%%%%%%%%%%%%%%%%%%%%%%%%%%%%%%%%%%%%%%%%%%%%%%
\newpage
\begin{quotation}
Table II. Numerical results for the special values of $b_\nu$.
The values with the underlines are input values.
The values with the parentheses denote those which contradict 
with the observed values. 
\end{quotation}
$$
\begin{array}{|c|c|c|}\hline\hline
b_\nu & -\exp[i \ 1.8^\circ] & -(1/2)\exp[i \ 0.12^\circ] \\ \hline
r \equiv\Delta m^2_{32}/\Delta m^2_{21} & 1.950 \times 10^4 & 0.03310 \\
\Delta m^2_{32} & 3.51 \times 10^2 \ {\rm eV}^2 
& \underline{0.016\ {\rm eV}^2} \\
\Delta m^2_{21} & \underline{0.018\ {\rm eV}^2} 
& 0.483 \ {\rm eV}^2 \\ \hline
m_0/\lambda & 37.5 \ {\rm eV} & 3.04 \ {\rm eV} \\ \hline
m_3^\nu & 18.7 \ {\rm eV} & 0.707 \ {\rm eV} \\
m_2^\nu & 0.176 \ {\rm eV} & 0.695 \ {\rm eV} \\
m_1^\nu & 0.114 \ {\rm eV} & 0.00164 \ {\rm eV} \\ \hline
\sum m_i^\nu & 19.0 \ {\rm eV} & 1.40 \ {\rm eV} \\ 
\langle m_\nu \rangle & 0.00705 \ {\rm eV} 
& 0.00267 \ {\rm eV} \\ \hline
P_{ee} & 0.523  & (0.990) \\
\sin^2 2\theta_{atm} & 0.900 & 0.995 \\
\sin^2 2\theta_{LSND} & (0.000048) & 0.0191 \\
\sin^2 2\theta^{\mu\tau} & 0.211 & 5.4\times 10^{-6} \\ \hline
m_0 & 3.12 \times 10^2 \ {\rm GeV} &  3.12 \times 10^2 \ {\rm GeV}\\
\kappa m_0 & 1.54 \times 10^{10} \ {\rm GeV} &  
1.90 \times 10^{11} \ {\rm GeV}\\
\lambda m_0 & 2.60 \times 10^{12} \ {\rm GeV} & 
 3.20 \times 10^{13} \ {\rm GeV}\\ \hline\hline
\end{array}
$$

\end{document}